\documentclass[twocolumn,tighten,table]{aastex631}
\usepackage{natbib}
\usepackage{amsmath}
\usepackage{listings}
\usepackage[normalem]{ulem}
\usepackage[section]{placeins}
\usepackage{color}
\usepackage{graphicx}
\usepackage{fancyvrb}
\usepackage{float}
\usepackage[T1]{fontenc}

\shorttitle{In-situ formation of primordial star clusters at $z > 7$ via gaseous disk fragmentation}
\shortauthors{L.~Mayer et al.}

\newcommand{\comments}[1]{} %usage: \comments{}

\newcommand{\soutPC}{\bgroup\markoverwith{\textcolor{cyan}{\rule[0.5ex]{2pt}{1pt}}}\ULon}
\newcommand{\soutFD}{\bgroup\markoverwith{\textcolor{magenta}{\rule[0.5ex]{2pt}{1pt}}}\ULon}

\newcommand\T{\rule{0pt}{2.6ex}}       % Top strut
\newcommand\B{\rule[-1.2ex]{0pt}{0pt}} % Bottom strut

\begin{document}

\title{In-situ formation of star clusters at $z > 7$ via galactic disk fragmentation; shedding light on ultra-compact
clusters and overmassive black holes seen by JWST}

\correspondingauthor{Lucio Mayer}
\email{lucio.mayer@uzh.ch}

\author[0000-0002-7078-2074]{Lucio Mayer}
\affiliation{Department of Astrophysics, University of Zurich, Winterthurerstrasse 190, CH-8057 Z\"urich, Switzerland}

\author[0000-0002-7235-9747]{Floor van Donkelaar}
\affiliation{Department of Astrophysics, University of Zurich, Winterthurerstrasse 190, CH-8057 Z\"urich, Switzerland}

\author[0000-0003-1427-2456]{Matteo Messa}
\affiliation{INAF -- OAS, Osservatorio di Astrofisica e Scienza dello Spazio di Bologna, via Gobetti 93/3, I-40129 Bologna, Italy}

\author[0000-0002-1786-963X]{Pedro R. Capelo}
\affiliation{Department of Astrophysics, University of Zurich, Winterthurerstrasse 190, CH-8057 Z\"urich, Switzerland}

\author[0000-0002-8192-8091]{Angela Adamo}
\affiliation{Department of Astronomy, Oskar Klein Centre, Stockholm University, AlbaNova University Centre, SE-106 91 Stockholm, Sweden}

% Abstract of the paper
\begin{abstract}
We investigate the nature of star formation in gas-rich galaxies at $z > 7$ forming in a markedly overdense region, in the whereabouts of a massive virialized halo already exceeding $10^{12}$~M$_{\sun}$. We find that not only the primary galaxy, but also the lower-mass companion galaxies rapidly develop massive self-gravitating compact gas disks, less than 500~pc in size, which undergo fragmentation by gravitational instability into very massive bound clumps. Star formation proceeds fast in the clumps, which quickly turn into compact star clusters with masses in the range $10^5$--$10^8$~M$_{\sun}$ and typical half-mass radii of a few pc, reaching characteristic densities above $10^5$~M$_{\sun}$~pc$^{-2}$.
The properties of the clusters in the lowest-mass galaxy bear a striking resemblance to those recently discovered by the James Webb Space Telescope (JWST) in the lensed Cosmic Gems arc system at $z = 10.2$. We argue that, due to their extremely high stellar densities, intermediate-mass black holes (IMBHs) would form rapidly inside the clusters, which would then swiftly sink and merge on their way to the galactic nucleus, easily growing a $10^7$~M$_{\sun}$  supermassive black hole (SMBH). Due to the high fractional mass contribution of clusters to the stellar mass of the galaxies, in the range 20--$40\%$, the 
central SMBH would comprise more than $10\%$ of the mass of its host galaxy, naturally explaining the overmassive SMBHs discovered by JWST at $z > 6$.
\end{abstract} %wordcount = 324


%%%%%%%%%%%%%%%%%%%%%%%%%%%%%%%%%%%%%%%%%%%%%%%%%%

%%%%%%%%%%%%%%%%% BODY OF PAPER %%%%%%%%%%%%%%%%%%

\section{Introduction}
The advent of the James Webb Space Telescope  {\citep[JWST, ][]{Gardner_et_al_2006}} is revolutionizing our view of the early stages of galaxy, black hole (BH), and star formation. Amongst the most startling discoveries concerning galaxies and stars are the unexpected abundance of disk-like galaxies at redshifts as high as 7--8 
{\citep[e.g.][]{Fujimoto_et_al_2024, Ferreira_et_al_2022,Ferreira_et_al_2023}
corroborated by Atacama Large Millimeter/submillimeter Array (ALMA) studies {\citep[]{Rowland_et_al_2024}}, the existence of already quenched galaxies at similarly high redshift \citep[][]{Looser_et_al_2024,Weibel_et_al_2024}, and, most recently, the existence of very dense compact star clusters at redshifts as high as 8-10, revealed owing to magnification via gravitational lensing \citep[e.g.][]{Adamo_et_al_2024,Mowla_et_al_2024}. Concurrently, a plethora of new detections of high-redshift supermassive BHs (SMBHs) has shown that many of these objects are overmassive with respect to their own host galaxies, when compared to local scaling relations, both at the high-mass \citep[e.g.][]{Stone_et_al_2024,Yue_et_al_2024} and low-mass \citep[e.g.][]{Harikane_et_al_2023, Maiolino_et_al_2023} end of the spectrum. The observed high-redshift compact clusters appear akin to clusters previously detected at somewhat lower redshift also in lensed sources, including the Sunburst and Sunrise arcs \citep[e.g.][]{Vanzella_et_al_2019,Pascale_et_al_2022,Vanzella_et_al_2023,Welch_et_al_2023}, and, while more compact, their properties hint to similarities with both globular clusters \citep[GCs;][]{Mowla_et_al_2022,Adamo_et_al_2023} and giant stellar clumps detected in star-forming galaxies at redshifts $<5$ \citep[e.g.][]{Guo_et_al_2015}. To explain the origin of giant clumps observed at cosmic noon in Hubble Space Telescope (HST) studies \citep[][]{Elmegreen_et_al_2009,Guo_et_al_2015,Shibuya_et_al_2016,Guo_et_al_2018,Zanella_et_al_2019}, the prevalent theory is gas disk fragmentation. Fragmentation can be understood via the \citet{Toomre_1964} instability theory augmented by the important role of mass loading via cold gas accretion, which prevents the system from reaching a well defined background equilibrium state \citep[sometimes called ``violent disk instability''; see, e.g.][]{Mandelker_et_al_2017}.

The disk mass and its gas fraction have been found to be key parameters controlling the susceptibility to fragmentation and  stellar clump formation \citep[e.g.][]{Tamburello_et_al_2015, Renaud_et_al_2021, vanDonkelaar_et_al_2022}, together with the effect of stellar and supernovae feedback \citep[][]{Mayer_et_al_2016,Oklopvcic_et_al_2017}. If disks are ubiquitous even at much higher redshift than previously expected, as suggested by JWST observations \citep[][]{Ferreira_et_al_2022,Ferreira_et_al_2023}, it is conceivable that fragmentation via disk instability should also happen at the redshifts accessible to JWST, especially because the cold gas content of galaxies  at such epochs should be comparable or higher than at $z = 1$--3.

\citet{Nakazato_et_al_2024} propose an alternative scenario for the origin of stellar clumps and clusters seen by JWST at $z > 6$, namely that they are produced during mergers of gas-rich galaxies. However, they reach this conclusion by studying a sample of simulated galaxies that span quite a narrow mass range at the final time, corresponding to $z = 5$, typically comprising galaxies as massive as the 
present-day Milky Way. Instead, some of the discovered systems of clusters at the highest redshifts, such as the Firefly Sparkler and Cosmic Gems \citep[][]{Adamo_et_al_2024,Mowla_et_al_2024} are hosted in relatively low mass galaxies. 

Here we present the results of the first high-redshift pc-resolution zoom-in galaxy formation simulation demonstrating the formation of massive star clusters in primordial disk galaxies at $z > 7$. These compact star clusters have properties reminiscent of those of the clusters recently discovered in JWST observations of lensed galaxies at $z = 7$--10. We argue that this is a common mode of star cluster formation in highly overdense regions associated with high-$\sigma$ peaks among the galaxy population at high redshift. 

\section{Numerical Simulations}
We use the very high-resolution cosmological hydrodynamical simulation MassiveBlackPS, originally carried out with the smoothed-particle hydrodynamics (SPH), $N$-body \textsc{Gasoline2} code \citep[][]{Wadsley_et_al_2017} to study the formation of massive BHs at high redshift via the direct ``dark collapse'' scenario \citep[][]{Mayer_et_al_2024}. The details of the complex numerical procedure developed to generate the simulation are extensively described in \citet{Mayer_et_al_2024}, where also the implementation of metallicity-dependent radiative cooling, cosmic ionizing background, and sub-grid physics models for star formation and supernovae feedback  are outlined. Here we just summarize the most relevant information. The parent simulation is the  large-scale cosmological hydrodynamical simulation MassiveBlack, with box size of {$\sim$}0.7~Gpc comoving \citep[][]{DiMatteo_et_al_2012,Feng_et_al_2014}. The initial conditions used in this paper are obtained after various re-sampling steps involving repeated application of the zoom-in technique on a selected sub-volume comprising a massive galaxy, which allows to reach a gas mass (dark matter) resolution of $1.9 \times 10^4$~M$_{\sun}$ ($9.4 \times 10^4$~M$_{\sun}$) and a gas (dark matter) gravitational softening of 142~pc (241~pc). Additionally, we applied particle splitting \citep[][]{Roskar_et_al_2015} of the gas phase  within the entire selected region. The latter corresponds to the virialized region of a massive galaxy undergoing a major merger. Due to the computational burden involved, the re-sampled simulation follows a timescale of only 60~Myr, from just before to just after the completion of the galaxy merger. In this work, we focus on a short evolutionary phase after the merger has terminated, at $z = 7.6$, lasting about 6 Myr. This stage of the simulation  has an exquisite resolution of about $2.4 \times 10^3$~M$_{\sun}$ in the gas phase, with the gravitational softening set to 2~pc. The resolution of hydrodynamical forces, given by the SPH smoothing length, is comparable or higher, depending on the local density, hence any result concerning gravitational collapse will be conservative \citep[][]{Bate_Burkert_1997}.

By construction, the targeted virialized region is that of a halo at the very high mass end of the galaxy mass function at $z > 6$, corresponding to a 4--5$\sigma$ peak
\citep[][]{Mayer_et_al_2024}. It is thus a highly overdense region, consistent with the original motivation for the selection of this specific halo in the MassiveBlack volume, namely its mass had to be consistent with the typical halo mass of hosts of the bright high-redshift quasars.  The virialized region contains also two galaxy companions, which we analyze in this Letter.

\section{Results}

\begin{figure*}
\centering
\setlength\tabcolsep{2pt}%
\begin{center}
\includegraphics[trim={0cm 0cm 0cm 0cm}, clip, width=0.92\textwidth, keepaspectratio]{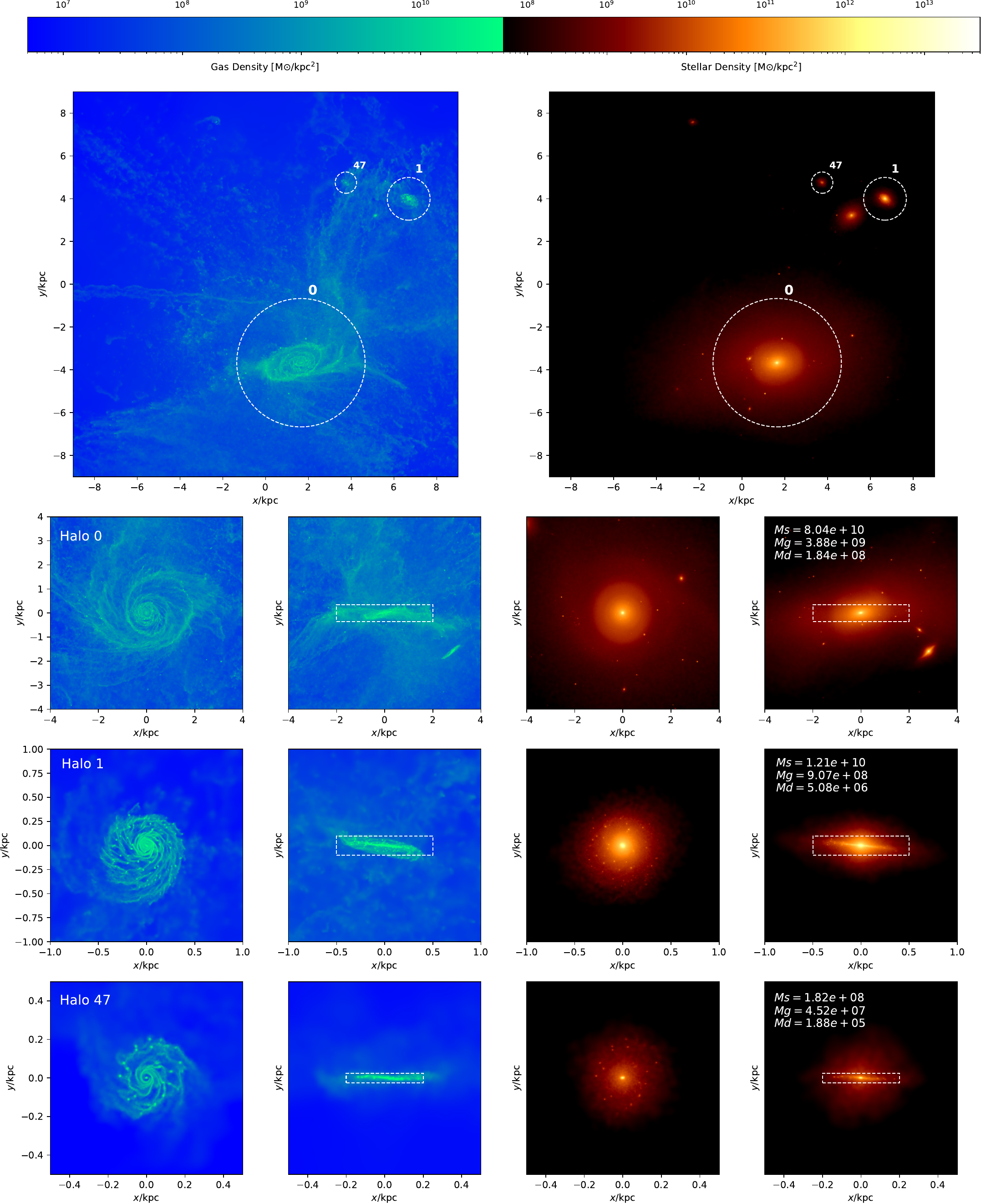}
\caption{Colour-coded density maps showing the location of the galaxy in the analyzed volume (top panels), the gas component of the disk, viewed face-on and side-on (left-hand panels), and the stellar distribution (right-hand panels) for, respectively, the primary galaxy (Halo 0), the intermediate-mass galaxy (Halo 1), and the lowest-mass galaxy (Halo 47), whose properties more akin to those in the Cosmic Gems system. The top panels show the sub-volume of radius about half the virial radius, which is the region within which particle splitting was applied \citep[see][]{Mayer_et_al_2024}. The rectangles in the side-on view represent the region of the disk where the stellar, gas, and dark matter masses (Ms, Mg, and Md, respectively) were calculated, with values presented in the rightmost panel in solar masses. The density maps depict the simulation approximately 5.6 Myr after the galaxy merger, near the end of the simulation analyzed in this paper. }
\label{fig:densitymaps}
\end{center}
\end{figure*}

The most massive, central galaxy in the simulation is a disk-dominated system with a  total stellar mass  of about $8 \times 10^{10}$~M$_{\sun}$ and size about 2~kpc (see Figure~\ref{fig:densitymaps}, top panels). It is the result of a major merger of two gas-rich disk galaxies, and its properties and
assembly are described in \citet{Mayer_et_al_2024}, where
we also show that it is consistent with the properties of
vigorously star forming galaxies above the main sequence at such
correspondingly high redshifts.
The dense gas disk undergoes copious fragmentation due to gravitational instability. 
Disk fragmentation takes place also in two companion galaxies located within the virial volume, whose baryonic masses are, respectively, about 8 times and 400 times lower than the mass of the primary galaxy (see Figure~\ref{fig:densitymaps}). These galaxies are extremely compact (250 to 500~pc in size) and host a proportionally more massive gas component in their disk compared to
the primary galaxies, close to half of the stellar mass in the
smallest galaxy (see Figure~\ref{fig:densitymaps}).
Disk fragmentation via gravitational instability is known to occur on a dynamical time-scale once favourable conditions are met. 
Due to the high densities in the disks
their dynamical time-scales  are short, of order of 1~Myr.
Heating
and stirring by supernova (SN) feedback is too slow to quench fragmentation by stabilizing the disk, a situation that bears some resemblance with
the feedback-free scenario for high-redshift massive galaxy formation
proposed by \citet{Dekel_et_al_2023}.
The products of fragmentation are massive and dense gravitationally bound gas clumps which rapidly form stars. By the end of the simulation, many have converted more than  $50\%$ of their mass in stars. At this point we consider them bona-fide star clusters. While our study
focuses on such star clusters, fragmentation of the gas disk continues until the end of our simulation, especially in the galaxy companions which are proportionally more  gas-rich, so that there is a non-negligible
population of clumps that are still dominated by gas at the end of the
simulation. Specifically, by the end nearly 30\% of them exceed a $70\%$ gas fraction in halo 47, whereas the same high gas fraction is reached
by less than $10\%$ of them in the primary galaxy, halo00 (middle panel, Figure \ref{fig:mass_histograms}).

\begin{figure}
\centering
\setlength\tabcolsep{2pt}%
\includegraphics[trim={0cm 0cm 0cm 0cm}, clip, width=0.45\textwidth, keepaspectratio]{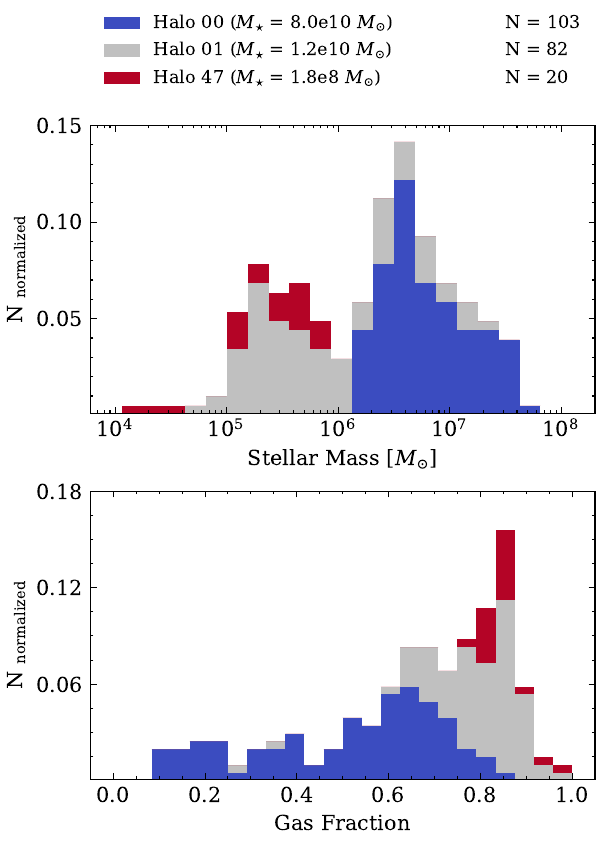}
\includegraphics[trim={0cm 0cm 0cm 0cm}, clip, width=0.45\textwidth, %keepaspectratio]{Ms_ClumpsEvol.pdf}
keepaspectratio]{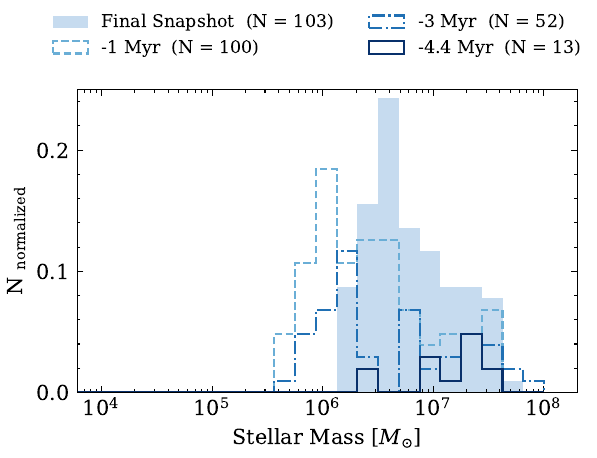}
\caption{Top panel: stellar mass histograms for the identified star clusters in the three galaxies, Halo00 being the primary galaxy, Halo01 the most massive galaxy companion, and Halo47 the lowest-mass galaxy companion. Middle panel: distribution of gas fractions of the star clusters in the three galaxies. Both of these panels are scaled such that the cumulative total of all star clusters equals 1. Bottom panel: time evolution of the mass distribution of stellar clumps/clusters in the primary galaxy, starting from the first snapshot from which the group finder first identifies gravitationally bound structures. The histograms in this panel are normalized so that the cumulative mass of star clusters in the final snapshot is 
set to 1.} 
\label{fig:mass_histograms}
\end{figure}

The top panel of Figure~\ref{fig:mass_histograms} shows the stellar mass function of the star clusters at the end of the simulation for all three galaxies\footnote{A central massive stellar nucleus, about 10 pc in size, forms in all the galaxies discussed, and has been excluded from the mass function because it is unrelated to star cluster formation via disk fragmentation.}.  The star clusters are identified via their gravitationally bound mass,
including both gas and stars. This corresponds to a characteristic radius
defining the bound region, and this is the reference radius which we use in the
following analysis.

We note that the star clusters in the disk span a wide range of masses, over more than three orders of magnitude, from comparable to low-mass GCs to much larger, with the higher mass end being akin to the mass of stellar clumps at $z = 2$--3 \citep[see, e.g.][]{Tamburello_et_al_2015,Tamburello_et_al_2017, Dessauges:2017aa}. For each galaxy we also compute the fractional mass contribution of the mass of the clusters relative to the total stellar mass of the disk excluding the central stellar nucleus. We find that the fractional mass contribution is in the range  $20-30\%$, hence a prominent fraction of the star formation occurs in this clustered mode.

In the main galaxy, as well as in the two companions, the Toomre parameter of the gas disk is lower than unity across a significant region in the disk, consistent with the onset of fragmentation. 
We remark that, when computing the Toomre parameter, we take into account that the gas disks have significant non-radial motions, resulting in an effective turbulent velocity which adds to the thermal sound speed \citep[see also][]{Romeo_Agertz_2014,Fiacconi_et_al_2017}.

At the onset of fragmentation, one can compare the characteristic size and mass of clumps  with the local value of the most unstable Toomre wavelength, $\lambda_{mu} = 2 {\pi}^2 G \Sigma_g /  {\kappa}^2$,
where $\Sigma_g$, $\kappa$ and $G$ are, respectively, the gas surface
density, the epicyclic frequency and the gravitational constant.
In order to compute the actual fragmentation scale, we adopt a modification of the latter equation which takes into account that fragmentation is a three-dimensional process taking place inside spiral arms \citep[][]{Boley_et_al_2010,Tamburello_et_al_2015}, rather than in the axisymmetric razor-thin disk assumed in the standard formulation of Toomre instability. This modification has been  thoroughly tested in past work, and typically
yields a length sale a factor 4--6  smaller
than the Toomre wavelength
because the relevant overdensity (at the denominator of the Toomre wavelength expression) is now that along the spiral arm, which is 
higher than in the  background axisymmetric flow \citep[][]{Boley_et_al_2010}.
We find  that the median most unstable Toomre wavelength across 
the gas disk ranges from 30 to 60 pc in the two companion galaxies (halo01 and halo47), so that the  actual fragmentation scale should be in the range 5 to 12 pc. The latter is in good agreement with the gravitationally bound radii of the clusters in the simulation, in the range 4 to 13 pc\footnote{We note that, in Figure \ref{fig:mass-density}, we present the half-mass radii of the clusters, which are 1.5–3 times smaller than their gravitationally bound radii. This scaling provides a more direct comparison with observations, which typically report half-mass radii derived from observed half-light radii.}. The same analysis is less straightforward in the primary galaxy, which is a disturbed merger remnant for which the background gas surface density is highly fluctuating.

The characteristic masses of the clusters appear to correlate with the mass of the host galaxy, as shown in the top panel of Figure \ref{fig:mass_histograms}. If one assumes that the characteristic surface density is proportional to the disk mass, the correlation can be understood by noting that the Toomre wavelength grows linearly with the gas surface density. A proportionality between gas surface density and disk mass is expected in models of disk formation inside cold dark matter halos \citep[e.g.][]{Mo_et_al_1998}.

\begin{figure}
\centering
\setlength\tabcolsep{2pt}%
\includegraphics[trim={0cm 0cm 0cm 0cm}, clip, width=0.45\textwidth, keepaspectratio]{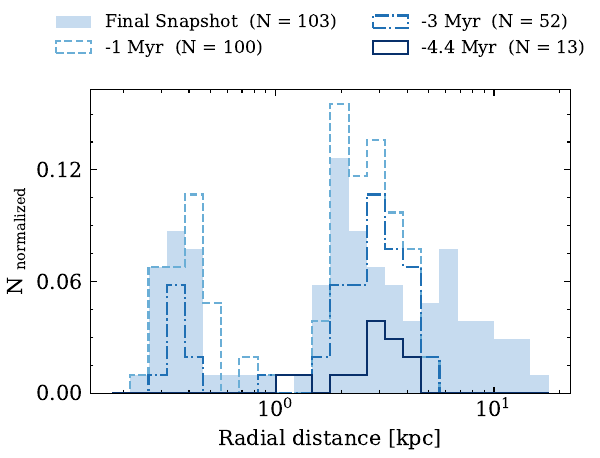}
\includegraphics[trim={0cm 0cm 0cm 0cm}, clip, width=0.45\textwidth, keepaspectratio]{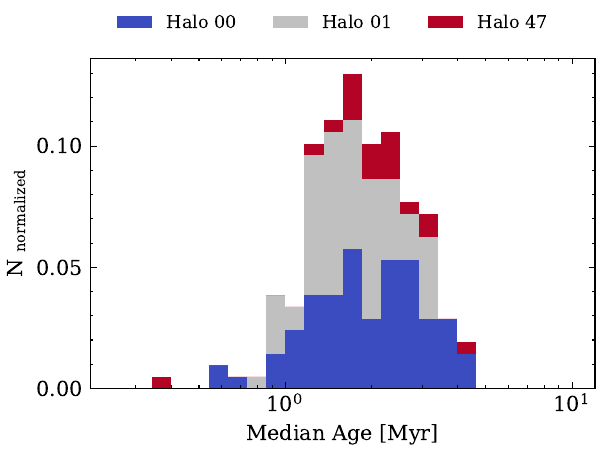}
\caption{Radial distribution (top panel) of the stellar clusters in halo 0 and mean age distribution (bottom panel) of clusters identified in all three galaxies at the end of the simulation. For the radial distribution, we show the evolution of star clusters in the massive galaxy only, halo 0, as it provides a much larger statistical sample. The repeated occurrence of peaks of the distribution in certain locations is related to dense ring-like overdensities in
which fragmentation occurs repeatedly. The timescales probed are not long enough to witness the expected smearing of such features by migration The normalisation of the histograms is the same as in Figure~\ref{fig:mass_histograms}.}
\label{fig:radial-age}
\end{figure}

We also study the time evolution of the clusters' stellar mass evolution of the primary galaxy (halo00, Figure~\ref{fig:mass_histograms}, bottom panel). We find rapid evolution, as expected based on the short dynamical time-scales, of the order of Myrs. In only 3~Myr the minimum mass shifts from a few $10^5$~M$_{\sun}$ to slightly above $10^6$~M$_{\sun}$, and the peak of the mass distribution also shifts to larger values as time progresses, which is a result of mergers between clusters as well as of mass accretion. At the same time, fragmentation continues over the course of the simulation, as shown by the increasing number of clusters as time progresses in e.g. the top panel of Figure~\ref{fig:radial-age}, and despite mergers acting to decrease their numbers. Yet, the increase in the number of clusters, in the primary galaxy, appears to slow down at the end.

The lower mass one among the two companion galaxies, halo47, is particularly interesting because its stellar mass (in the disk) is only a factor of 3--4 larger than that of the host galaxy of the Cosmic Gems arc recently studied by JWST at z=10.2. This has revealed a system of 5--6 ultra-compact, $\sim1$ pc-scale star clusters with masses up to a million solar masses \citep[][]{Adamo_et_al_2024}. The clusters are typically separated by a few tens of pc in projection, which is similar to the separation of the clusters in our simulation.  In the observations, the clusters span a range of ages in the few to 80~Myr range \citep[][]{Adamo_et_al_2024}. The ages of the star clusters in the simulated galaxies are thus consistent with those of the youngest stars of the  clusters of Cosmic Gems (Figure~\ref{fig:radial-age}, bottom panel), which is expected given the short duration of the simulation following
the onset of fragmentation.

\begin{figure}
\centering
\setlength\tabcolsep{2pt}%
\includegraphics[trim={0cm 0cm 0cm 0cm}, clip, width=0.50\textwidth, keepaspectratio]{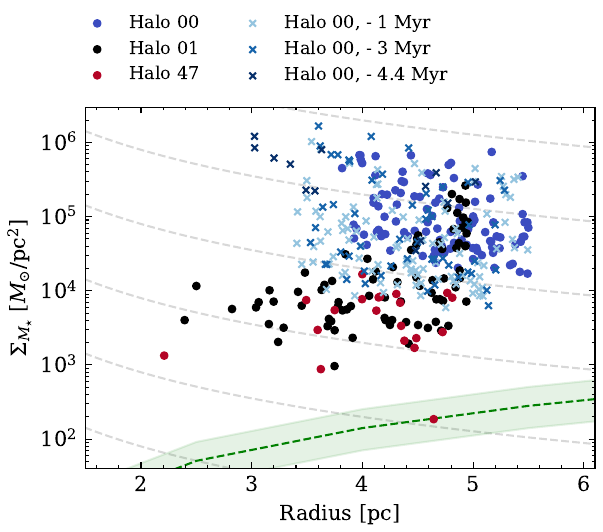}
\caption{Correlation between stellar surface density, $\Sigma_{M_{\star}}$,  and  half-mass radius of clusters stacking results for all three galaxies at the final time, and including also the results at different times for the primary galaxy. Gray dashed lines represent star clusters of similar mass, while the green line shows the stellar density-radius relation for clusters in local galaxies, following the best fit model from \citet{Brown_Gnedin_2021} using clusters from the LEGUS survey.}
\label{fig:mass-density}
\end{figure}

\begin{figure}
\centering
\setlength\tabcolsep{2pt}%
\includegraphics[trim={0cm 0cm 0cm 0cm}, clip, width=0.50\textwidth, keepaspectratio]{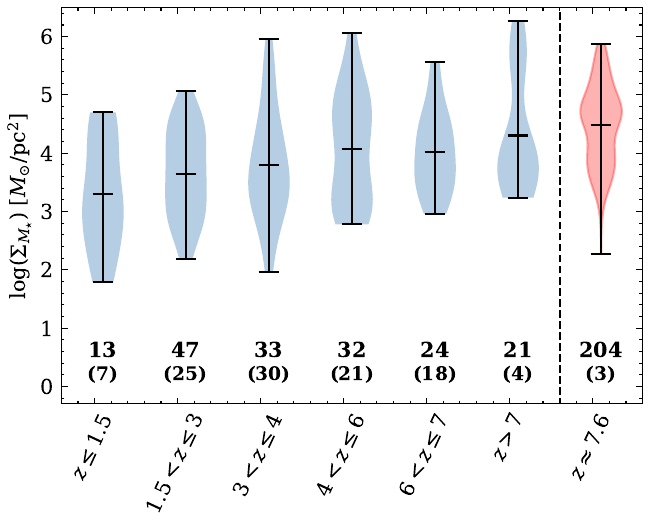}
\caption{Violin plots depicting the distribution of $\Sigma M_{\star}$ across different redshift intervals of stellar clusters with a R$_{\rm eff}$ below 25 pc. The numbers below the plot represent the number of clusters, with the number of galaxies indicated in parentheses. The blue violins correspond to observed data sets using JWST adapted from \citet{  Vanzella_et_al_2022b, Vanzella_et_al_2022, Vanzella_et_al_2023, Messa_et_al_2024b, Mestric_et_al_2022, Claeyssens_et_al_2023, Adamo_et_al_2024, Fujimoto_et_al_2024,  Mowla_et_al_2024, Claeyssens_et_al_2024}; Messa et al. (in prep.). In contrast, the red violin corresponds to the dataset from this study, where the surface density is computed at the half-mass radius as typically done in the
observations.}
\label{fig:MassDens_obs}
\end{figure}

We now turn to the structural properties of the star clusters. We show the relation between the stellar surface density and the half-mass radii in Figure~\ref{fig:mass-density}, and we compare the range of
stellar surface densities in the
simulated clusters with those of the
clusters discovered by JWST in Figure \ref{fig:MassDens_obs}. 
The stellar surface densities are computed at the half-mass radius. Overall, our results match the observations very well, especially at comparable redshifts. We note that the stellar surface densities are extremely high, 1-3 orders of magnitude higher than those of present-day star clusters \citep{Brown_Gnedin_2021}, on average. The clusters in this work span a wide range of surface densities, which includes several clusters above $10^5$~M$_{\sun}$~pc$^{-2}$, comparable to the clusters in the Cosmic Gems system. The reported densities are conservative because inspection of  the steep surface density profiles  of some of the clusters reveals that their central values can increase by
a further 2-3 orders of magnitude, although such inner regions can be below the gravitational softening length for the smaller clusters and thus will not be discussed further.
 
Given the high density of the background galaxy and the initial small galactocentric separation at which the clusters form, one expects the clusters to migrate fast inward by dynamical friction. Migration will be accompanied by gas inflows, and together they contribute to increase the mass of the central region and assemble a bulge or dense nucleus \citep[][]{Tamburello_et_al_2017, vanDonkelaar_et_al_2024a}. The imprint of migration on the radial distribution of clumps at different times is not evident, though, instead persistent 
peaks are present in the radial distribution  (see Figure~\ref{fig:radial-age}). One reason is that the expected migration timescale to the galactic center, estimated using the dynamical friction timescale via \citeauthor{chandra:1943aa}'s formula, is $> 10^7$
yr for most of the clusters (see Section \ref{sec:disc}), hence longer than the time span
of the simulations. Additionally, clump-clump interactions and new episodes of fragmentation play an important role (Figure~\ref{fig:radial-age}, top panel). The age distribution (Figure~\ref{fig:radial-age}, bottom panel) also shows that fragmentation  is continuously going on as some clusters with ages less than 1~Myr are present even at the end of the simulation.

Due to the phenomenally high central stellar densities found in the star clusters, it is conceivable that an intermediate-mass black hole (IMBH) might form at their center. \citet{Fujii_et_al_2024} have carried out the first star-by-star simulation of GC progenitors, showing that, with densities in the range $10^3$--$10^4$~M$_{\sun}$~pc$^{-3}$ at pc scales, the formation of very massive stars is promoted as mass loss via stellar winds is superseded by stellar mergers, resulting in IMBHs with final masses exceeding $10^3$~M$_{\sun}$. In our clusters, the densities at pc scales are at a minimum comparable and often 2--3 orders of magnitude larger than those in the proto-clusters of \citet{Fujii_et_al_2024}, suggesting that IMBH formation should be the norm in these systems (see also \citet{Rantala_et_al_2024}). Furthermore, \citet{Fujii_et_al_2024} find that the mass of the IMBH should scale as the square  root of the mass of the proto-clusters. Assuming the latter scaling, in the most massive of our clusters, which exceeds $10^9$~M$_{\sun}$, IMBHs with masses up to $10^5$~M$_{\sun}$ should arise. 

\section{Discussion and Conclusion} \label{sec:disc}

Our disk fragmentation scenario explains naturally the origin of the massive high-redshift compact clusters recently discovered by JWST. It exhibits two distinctive features: (i) a rapid  formation time-scale, which is fairly homogeneous across the entire galaxy, since gravitational instability occurs on a time-scale proportional to the dynamical time in the galaxy's disk, and (ii) a dominant contribution to star formation due to its ability to convert gas into stars more efficiently in the  clumps than in the background gas disk. These features result in a narrow age spread of the clusters and a high fractional mass contribution to the stellar mass of the galaxy, both being consistent with the clusters discovered by JWST \citep[][]{Adamo_et_al_2024}.

Using the standard \citeauthor{chandra:1943aa}'s formula yields that the dynamical friction time-scale of the clusters is $<10^8$~yr, even for the clusters at the low-mass end of the mass distribution shown in Figure~\ref{fig:mass_histograms}, with the most massive clusters sinking in roughly $10^7$~yr. If there is a central MBH seed at the center of the galaxy, the clusters will thus merge with it. If the central BH has a mass of $10^4$--$10^5$~M$_{\sun}$, as in the most conventional variants of the direct collapse scenario, cluster merging with the central BH would be a major growth channel, competitive with gas accretion, because even in the smallest of the three galaxies the total mass in clusters exceeds $10^7$~M$_{\sun}$ \citep[see also][]{vanDonkelaar_et_al_2024b}. The accreted mass would come in the form of stars, residual gas as well as compact objects, releasing a wealth of tidal disruption events. The
inward mass flux is very high, being of order $10^7$~M$_{\sun}$$/10^8$~yr~$= 0.1$~M$_{\sun}$~yr$^{-1}$ if we choose, conservatively, the upper
limit for the dynamical friction timescale.

However, as we outlined above, a second scenario is possible, namely it is likely that IMBHs with masses in the range  $10^3$--$10^5$~M$_{\sun}$ would arise in the clusters, likely over a few central crossing times, i.e. over $<10^5$ yr, which is much less than the aforementioned dynamical friction time-scale. The IMBHs would also sink to the center via dynamical friction and assemble a supermassive black hole (SMBH). This second scenario is conceptually
similar to the model proposed by \citet{Dekel:2024aa} in feedback-free galaxies
(see also \citet{Lupi_et_al_2016})

The key conclusion is that, either by means of direct accretion of ultra-compact clusters on a pre-existing small BH seed, or
via the merging of IMBHs formed at the
center of the star clusters, a central SMBH with a mass of at least $10^7$~M$_{\sun}$
can be assembled. The resulting SMBH would not be as large as those powering the bright, high-redshift quasars, but its mass would be comparable to the SMBHs discovered by JWST \citep[e.g.][]{Harikane_et_al_2023,Larson_et_al_2023,Maiolino_et_al_2023}, and also of the same order
of that predicted by Super-Eddington gas accretion models on light BH seeds inside proto-galaxies \citep[e.g.][]{Sassano_et_al_2023}.

The SMBH formation scenario that we have just outlined would occur in all the galaxies studied in this paper. Interestingly, for a small galaxy such as halo47, the outcome would be an SMBH with
a mass only a factor of a few smaller than
the stellar mass of the host.
This ``IMBH rain scenario'' can thus naturally explain the recent JWST observations of overmassive SMBHs \citep[e.g.][]{Harikane_et_al_2023,Maiolino_et_al_2023,Stone_et_al_2024,Yue_et_al_2024}. 

In the more massive primary galaxy, a pre-existing central SMBH already weighting  $10^7$~M$_{\sun}$ could arise via  "dark collapse"  \citep[][]{Zwick_et_al_2023},
whereby a supermassive proto-star contracts
by the General Relativistic radial instability forming directly a SMBH.
In this case, the main growth mechanism
would be  gas accretion through a
a central gaseous 
supermassive disk (SMD),
as shown in \citet{Mayer_et_al_2024},
eventually resulting in a billion solar mass SMBH  in much less than 1 Gyr \citep[][]{Zwick_et_al_2023}.
%In this case
%the fractional mass growth would be comparatively smaller (owing to compact object mergers, including IMBHs and TDEs), but still non-negligible.  

In summary, we argue that, in highly biased virialized patches at high
redshift, two different formation mechanisms
would dominate the assembly and growth of SMBHs, depending on the mass of the galaxy; repeated mergers of ultra-compact
clusters and IMBHs in low to average mass galaxies, leading to 
SMBHs with moderate mass which are overmassive relative to the host
galaxy, and dark collapse in the most massive galaxies, leading to the most massive SMBHs that power
the brightest high-redshift Quasars.
In our scenario, the occurrence of overmassive SMBHs should thus be common
since it can happen in the most abundant
population of galaxies.

We have reported that clusters have still significant amounts of
gas by the end of the simulations.
However, in the sub-grid treatment of our simulations  radiative feedback from massive 
stars is not included. The latter
would heat and ionize gas inside clusters already on the Myr timescales that we
are following here
would test our formation scenario.
The gas fractions of the clusters at the end of the simulation might thus be overestimated. On the other end, the escape speeds from our clusters are typically larger than those in the million solar mass clouds simulated by \citet{Dale_et_al_2012}, in which the effect of ionizing stellar radiation was found to be too weak to expel significant amounts of gas. 
Assessing whether or not  a cold gas component is present in the clusters detected by JWST, perhaps with ALMA, would test our scenario further.

In this paper we studied a highly biased, overdense virialized region in which one expects galaxies to grow faster than average. The rapid baryonic mass growth 
should promote gas disk fragmentation
relative to less biased regions.
To address qualitatively the role of the environment, we turn to a different high-res simulation, the GigaEris simulation \citep[][]{Tamfal_et_al_2022}, which has comparable resolution, same sub-grid physics, but follows a significantly lower-mass virialized region (at $z=7$ the virial mass of the primary halo is nearly two orders of magnitude lower than that of the halo selected MassiveBlackPS),   which results into to a typical star-forming galaxy at $z = 5$--6. Neither the main galaxy nor the companion galaxies undergo any sign of disk fragmentation \citep[][]{vanDonkelaar_et_al_2023}.
Indeed the mass growth rate of the disk of the galaxy is much
lower than in the primary galaxy of MassiveBlackPS, with its stellar
mass at $z=5$ being still about ten times lower than that of our primary
at $z = 7.6$.
Future observations of the larger scale environments of the host galaxies
of  the massive clusters in the JWST observation would provide an additional test of our scenario by assessing if disk fragmentation, in both low and high
mass galaxies, occurs predominantly in highly overdense regions.

\begin{acknowledgements}
LM and FvD acknowledge support from the Swiss National Science Foundation under the Grant 200020\_207406. MM acknowledges the financial support through grant PRIN-MIUR 2020SKSTHZ. PRC acknowledges support from the Swiss National Science Foundation under the Sinergia Grant CRSII5\_213497 (GW-Learn). AA acknowledges support by the Swedish research council Vetenskapsradet (2021-05559).
\end{acknowledgements}

\bibliographystyle{aasjournal}
\bibliography{output}

\end{document}